\documentclass[twocolumn,aps,prc,showpacs,superscriptaddress,nofootinbib,floatfix]{revtex4}
\usepackage{graphicx}
\usepackage{dcolumn}
\usepackage{bm}
\usepackage{longtable}

\begin{document}
\def\Journal#1#2#3#4{{#1} {\bf #2}, #3 (#4)}

\def\NCA{Nuovo Cimento}
\def\NIM{Nucl. Instr. Meth.}
\def\NIMA{{Nucl. Instr. Meth.} A}
\def\NPB{{Nucl. Phys.} B}
\def\NPA{{Nucl. Phys.} A}
\def\PLB{{Phys. Lett.}  B}
\def\PRL{Phys. Rev. Lett.}
\def\PRC{{Phys. Rev.} C}
\def\PRD{{Phys. Rev.} D}
\def\ZPC{{Z. Phys.} C}
\def\JPG{{J. Phys.} G}
\def\CPC{Comput. Phys. Commun.}
\def\EPJ{{Eur. Phys. J.} C}

\topmargin 0pt
\title{Growth of Long Range Forward-Backward Multiplicity Correlations with Centrality in Au+Au Collisions at $\sqrt{s_{NN}}$ = 200 GeV }
\affiliation{Argonne National Laboratory, Argonne, Illinois 60439, USA}
\affiliation{University of Birmingham, Birmingham, United Kingdom}
\affiliation{Brookhaven National Laboratory, Upton, New York 11973, USA}
\affiliation{University of California, Berkeley, California 94720, USA}
\affiliation{University of California, Davis, California 95616, USA}
\affiliation{University of California, Los Angeles, California 90095, USA}
\affiliation{Universidade Estadual de Campinas, Sao Paulo, Brazil}
\affiliation{University of Illinois at Chicago, Chicago, Illinois 60607, USA}
\affiliation{Creighton University, Omaha, Nebraska 68178, USA}
\affiliation{Nuclear Physics Institute AS CR, 250 68 \v{R}e\v{z}/Prague, Czech Republic}
\affiliation{Laboratory for High Energy (JINR), Dubna, Russia}
\affiliation{Particle Physics Laboratory (JINR), Dubna, Russia}
\affiliation{Institute of Physics, Bhubaneswar 751005, India}
\affiliation{Indian Institute of Technology, Mumbai, India}
\affiliation{Indiana University, Bloomington, Indiana 47408, USA}
\affiliation{Institut de Recherches Subatomiques, Strasbourg, France}
\affiliation{University of Jammu, Jammu 180001, India}
\affiliation{Kent State University, Kent, Ohio 44242, USA}
\affiliation{University of Kentucky, Lexington, Kentucky, 40506-0055, USA}
\affiliation{Institute of Modern Physics, Lanzhou, China}
\affiliation{Lawrence Berkeley National Laboratory, Berkeley, California 94720, USA}
\affiliation{Massachusetts Institute of Technology, Cambridge, MA 02139-4307, USA}
\affiliation{Max-Planck-Institut f\"ur Physik, Munich, Germany}
\affiliation{Michigan State University, East Lansing, Michigan 48824, USA}
\affiliation{Moscow Engineering Physics Institute, Moscow Russia}
\affiliation{City College of New York, New York City, New York 10031, USA}
\affiliation{NIKHEF and Utrecht University, Amsterdam, The Netherlands}
\affiliation{Ohio State University, Columbus, Ohio 43210, USA}
\affiliation{Old Dominion University, Norfolk, VA, 23529, USA}
\affiliation{Panjab University, Chandigarh 160014, India}
\affiliation{Pennsylvania State University, University Park, Pennsylvania 16802, USA}
\affiliation{Institute of High Energy Physics, Protvino, Russia}
\affiliation{Purdue University, West Lafayette, Indiana 47907, USA}
\affiliation{Pusan National University, Pusan, Republic of Korea}
\affiliation{University of Rajasthan, Jaipur 302004, India}
\affiliation{Rice University, Houston, Texas 77251, USA}
\affiliation{Universidade de Sao Paulo, Sao Paulo, Brazil}
\affiliation{University of Science \& Technology of China, Hefei 230026, China}
\affiliation{Shandong University, Jinan, Shandong 250100, China}
\affiliation{Shanghai Institute of Applied Physics, Shanghai 201800, China}
\affiliation{SUBATECH, Nantes, France}
\affiliation{Texas A\&M University, College Station, Texas 77843, USA}
\affiliation{University of Texas, Austin, Texas 78712, USA}
\affiliation{Tsinghua University, Beijing 100084, China}
\affiliation{United States Naval Academy, Annapolis, MD 21402, USA}
\affiliation{Valparaiso University, Valparaiso, Indiana 46383, USA}
\affiliation{Variable Energy Cyclotron Centre, Kolkata 700064, India}
\affiliation{Warsaw University of Technology, Warsaw, Poland}
\affiliation{University of Washington, Seattle, Washington 98195, USA}
\affiliation{Wayne State University, Detroit, Michigan 48201, USA}
\affiliation{Institute of Particle Physics, CCNU (HZNU), Wuhan 430079, China}
\affiliation{Yale University, New Haven, Connecticut 06520, USA}
\affiliation{University of Zagreb, Zagreb, HR-10002, Croatia}
\author{B.~I.~Abelev}\affiliation{University of Illinois at Chicago, Chicago, Illinois 60607, USA}
\author{M.~M.~Aggarwal}\affiliation{Panjab University, Chandigarh 160014, India}
\author{Z.~Ahammed}\affiliation{Variable Energy Cyclotron Centre, Kolkata 700064, India}
\author{B.~D.~Anderson}\affiliation{Kent State University, Kent, Ohio 44242, USA}
\author{D.~Arkhipkin}\affiliation{Particle Physics Laboratory (JINR), Dubna, Russia}
\author{G.~S.~Averichev}\affiliation{Laboratory for High Energy (JINR), Dubna, Russia}
\author{J.~Balewski}\affiliation{Massachusetts Institute of Technology, Cambridge, MA 02139-4307, USA}
\author{O.~Barannikova}\affiliation{University of Illinois at Chicago, Chicago, Illinois 60607, USA}
\author{L.~S.~Barnby}\affiliation{University of Birmingham, Birmingham, United Kingdom}
\author{J.~Baudot}\affiliation{Institut de Recherches Subatomiques, Strasbourg, France}
% ...

\collaboration{STAR Collaboration}\noaffiliation 
\date{\today}
          
\begin{abstract}
Forward-backward multiplicity correlation strengths have been measured with the STAR detector for Au+Au and $\textit{p+p}$ collisions at $\sqrt{s_{NN}}$ = 200 GeV. Strong short and long range correlations (LRC)  are seen in central Au+Au collisions. 
The magnitude of these correlations decrease with decreasing centrality until only short range correlations are observed in peripheral Au+Au collisions. Both the Dual Parton Model (DPM) and the Color Glass Condensate (CGC) predict the existence of the long range correlations. 
In the DPM the fluctuation in the number of elementary (parton) inelastic collisions produces the LRC. In the CGC longitudinal color flux tubes generate the LRC. The data is in qualitative agreement with the predictions from the DPM 
 and indicates the presence of multiple parton interactions.
\end{abstract}
\maketitle

\newpage
The study of correlations among particles produced in different rapidity
regions may provide an understanding of the elementary (partonic) interactions 
which lead to hadronization.
Many experiments show strong short-range correlations (SRC) over a region of $\sim \pm $ 1 units in rapidity \cite{alner,aexopoulos}. 
In high-energy nucleon-nucleon collisions ($\sqrt{s} \gg$ 100 GeV) the 
nonsingly diffractive inelastic cross section increases significantly with energy and the magnitude of the long-range forward-backward multiplicity correlations (LRC) increases with the energy \cite{aexopoulos}. These effects can be understood in terms of multiparton interactions \cite{walker2004}.

 In high energy nucleus-nucleus collisions, it has been predicted that multiple parton interactions would produce larger long-range forward-backward multiplicity correlations that extend beyond $\pm$ 1 units in rapidity, compared to hadron-hadron scattering at the same energy \cite{capela1,capela2,larry}. The model based on multipomeron exchanges\\(Dual Parton Model) predicts the existence of long range correlations \cite{capela1,capela2}.
 In the Color Glass Condensate\\(CGC) picture of particle production the correlations of the particles created at early stages of the collisions can spread over large rapidity intervals, unlike the particles produced at later stages. Thus the measurement of the long range rapidity correlations of the produced particle multiplicities could give us some insight into the space-time dynamics of the early stages of the collisions \cite{larry}.  

One method to study the LRC strength 
is to measure the magnitude of the forward-backward multiplicity correlation over a long range in pseudorapidity. Such correlations were studied in several experiments \cite{alner,aexopoulos,na35,e802,tom,phobos,phenix} and investigated 
theoretically \cite{capela2,amelin,braun,larry,giov,shi,urqmd}.
In this paper we present the first results on the forward-backward (FB) multiplicity correlation strength and its range in pseudorapidity in heavy ion collisions at the Relativistic Heavy Ion Collider (RHIC) measured by the STAR experiment. 
Earlier analyses in STAR have focused on the relative correlations of charged particle pairs on the difference variables $\Delta\eta$ (pseudorapidity) and $\Delta\phi$ (azimuth). It was observed that the near-side peak is elongated in $\Delta\eta$ in central Au+Au as compared to peripheral collisions \cite{tom}. In the present work the measure of correlation strength as defined in Eq. (1) and the coordinate system differs from that of these earlier STAR measurements. The FB correlation strength is measured in an absolute
coordinate system, where $\eta $ = 0 is always physically located at
midrapidity (the collision vertex), instead of the relative $\eta $ difference
utilized in other 2-particle analyses. These differences allow the determination of the absolute magnitude of the correlation strength.

The correlation strength is defined by the dependence of the average charged particle multiplicity in the backward hemisphere, $\langle N_{b} \rangle$, on the
event multiplicity in the forward hemisphere, $N_{f}$, such that $\langle N_{b}\rangle = a + b N_{f}$, where $\textit{a}$ is a constant and $\textit{b}$ measures the correlation strength:  
\begin{equation}
 b = \frac{\langle N_{f}N_{b}\rangle - \langle N_{f}\rangle \langle N_{b}\rangle}{\langle N_{f}^{2}\rangle - \langle N_{f} \rangle ^{2}}= \frac{D_{bf}^{2}}{D_{ff}^{2}} 
\label{b}
\end{equation}  
In Eq. (1), $D_{bf}^{2}$(covariance) and $D_{ff}^{2}$(variance) are the backward-forward and forward-forward dispersions, respectively \cite{capela1,capela2}.

The data utilized for this analysis are from year 2001 (Run II) $\sqrt{s_{NN}}$ = 200 GeV minimum bias Au+Au collisions ($\sim$ 2.5$\times 10^{6}$ events) at the RHIC, as measured by the STAR experiment \cite{starnim}.  The FB correlation has been studied as a function of the centrality of the collision. 
 The centralities studied in this analysis account for 0-10, 10-20, 20-30, 30-40, 40-50 and 50-80\% of the total hadronic cross section. All primary tracks with distance of closest approach to the primary event vertex $< 3$ cm in the Time Projection Chamber (TPC) pseudorapidity range $|\eta|<1.0$ and with transverse momentum $p_{T} >$ 0.15 GeV/c were considered. This region was subdivided into bins of width $\eta = 0.2$. The FB intervals were located symmetrically about midrapidity ($\eta = 0$) with the distance between bin centers (pseudorapidity gap $\Delta\eta$): 0.2, 0.4, 0.6, 0.8, 1.0, 1.2, 1.4, 1.6, and 1.8. 
To avoid a bias in the FB correlation measurements, care was taken to use different pseudo-rapidity selections for the centrality determination which is also based on multiplicity.
Therefore, the centrality determination for the
 FB correlation strength for $\Delta\eta$ = 0.2, 0.4 and 0.6 is based on the multiplicity in  $0.5<|\eta|<1.0$, while for $\Delta\eta$ = 1.2, 1.4, 1.6 and 1.8 the centrality is obtained from $|\eta|<0.5$. For $\Delta\eta$ = 0.8 and 1.0 the sum of multiplicities from $|\eta|<0.3 $ and $0.8<|\eta|<1.0$ is used for the centrality determination.
The effect of centrality region selection on FB correlation strength was also studied by narrowing the region to $|\eta|<$ 0.3, 0.2 and 0.1 for all $\Delta\eta$ bins. This increases the FB correlation strength by $\sim$ 10-15\% at the most.
 Since the pseudorapidity particle density ($dN/d\eta$) plateau at $\sqrt{s_{NN}}$ = 200 GeV in Au+Au collisions extends over the region of interest \cite{phobos2}, this procedure yields consistent particle counts in the FB measurement intervals. An analysis of the data from (Run II) $\textit{p+p}$ collisions at $\sqrt{s}$ = 200 GeV, was also performed on minimum bias events ($\sim$ 3.5$\times 10^{6}$ events) using the same track cuts as for the Au+Au analysis.
 Corrections for detector geometric acceptance and tracking efficiency were carried out using a Monte Carlo event generator (HIJING) and propagating the simulated particles through a GEANT representation of the STAR detector geometry.

In order to eliminate (or at least reduce) the effect of impact parameter (centrality) fluctuations on the measurement of the FB correlation strength, each relevant quantity ($N_{f}$, $N_{b}$, $N_{f}^{2}$,  $N_{f}N_{b}$) was obtained on an event-by-event basis as a function of the event multiplicity, $N_{ch}$. The average uncorrected mean multiplicities
$\langle N_{f} \rangle_{uncorr}$, $\langle N_{b} \rangle_{uncorr}$, $\langle N_{f}^{2} \rangle_{uncorr}$, and $\langle N_{f}N_{b} \rangle_{uncorr}$ in each centrality bin were calculated from a fit to the  $N_{ch}$ dependences \cite{ebye1,tjtwinter}. The functional forms that were used are linear in $N_{f}$, $N_{b}$ and quadratic in $N_{f}^{2}$ and $N_{f}N_{b}$ for all $\Delta\eta$ bins. 
Tracking efficiency and acceptance corrections were then applied to $\langle N_{f} \rangle_{uncorr}$, $\langle N_{b} \rangle_{uncorr}$, $\langle N_{f}^{2} \rangle_{uncorr}$, and $\langle N_{f}N_{b} \rangle_{uncorr}$ to each event. Then the  corrected values of $\langle N_{f} \rangle$, $\langle N_{b} \rangle$, $\langle N_{f}^{2} \rangle$, and $\langle N_{f}N_{b} \rangle $ for each event were used to calculate the backward-forward and forward-forward dispersions, $D_{bf}^{2}$ and $D_{ff}^{2}$, binned by centrality, and normalized by the total number of events in each bin. This method removes the dependence of the FB correlation strength on the width of the centrality bin.
 As a cross check an alternative method of centrality determination was also carried out using the STAR Zero Degree Calorimeter (ZDC) for the 0-10\% centrality range and the results
are shown in Fig. 1a along with the 0-10\% most central events from the minimum bias dataset.
\begin{figure}[thbp]
\centering        
\vspace*{-0.2cm}
\includegraphics[width=0.56\textwidth,height=4.0in]{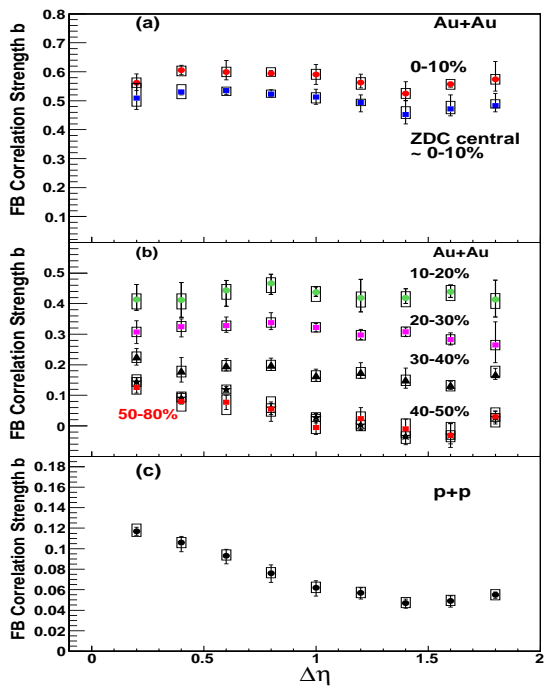}
\vspace*{-1.5cm}
\caption{(a) FB correlation strength for 0-10\% (circle) and ZDC based centrality ( square) (b) FB correlation strength for 10-20, 20-30, 30-40, 40-50 and 50-80\% ( square) Au+Au and (c) for $\textit {p+p}$ collisions as a function of $\Delta\eta$ at $\sqrt{s_{NN}}$ = 200 GeV. The error bars represent the systematic point-to-point error. The boxes show the correlated systematic errors.}
\label{CentralAu}
\end{figure}
  Statistical errors are smaller than the data points. 
Systematic effects dominate the error determination. The systematic errors are determined by binning events according to the  z-vertex in steps of 10 cm varying from -30 to 30 cm and the maximum value of the fit range (0-570, 0-600 and 0-630) for $\langle N_{f} \rangle, \langle N_{b} \rangle, \langle N_{f}^2 \rangle$, and $\langle N_{f}N_{b} \rangle$ vs $N_{ch}$. An additional error could arise due to finite detection efficiency in the TPC. This is estimated to be $\sim$ 5\% for most central collisions. 
 The overall systematic errors due to the fit range, which causes a
correlated shift along the y-axis, are shown in figures as boxes.

 Figure \ref{CentralAu} shows the FB correlation strength as a function of  $\Delta\eta$ for $\textit{p+p}$ and centrality selected Au+Au collisions along with the  ZDC based centrality results. The results from ZDC are slightly lower as compared to the 0-10\% most central events
 sampled from  minimum bias datasets using $N_{ch}$. 
It is observed that the magnitude of the  FB correlation strength decreases with the decrease in centrality. 
 The FB correlation strength evolves from a nearly flat function for 0-10\% to a sharply  decreasing function with $\Delta\eta$ for the 40-50 and 50-80\% centrality bins, which is expected if only short range correlations (SRC) are present \cite{capela1}.
The FB correlation strength values for 40-50 and 50-80\% centrality bins at large gap 
($\Delta\eta >$ 1.0) have an average value near zero. The individual b values are near zero within their systematic errors.
 Figure \ref{CentralAu} shows that the dependence of the FB correlation strength with $\Delta\eta$ is quite different in central Au+Au compared to $\textit{p+p}$ collisions. It is also observed that the FB correlation strength decreases faster in the peripheral (40-50 \% centrality) Au+Au as compared to $\textit{p+p}$ collisions. This indicates that the short range correlation length is smaller in 
Au+Au collisions than in $\textit{p+p}$.  

  Figure 2 shows the dependence of the dispersions $D_{bf}^{2}$ and $D_{ff}^{2}$ on $\Delta\eta$ for central Au+Au collisions (Fig. 2a) and $\textit{p+p}$ collisions (Fig. 2b). 
 The nearly constant value of $D_{ff}^{2}$ with $\Delta\eta$ represents the dispersion within the same $\eta$ window, which has approximately the same average multiplicity for all $\Delta\eta$ values. The behavior of  $D_{bf}^{2}$ is similar to the FB correlation strength. Thus the FB correlation variation with the size of $\Delta\eta$ is dominated by the $D_{bf}^{2}$ in Eq. (1).

\begin{figure}[thbp]
\vspace*{-0.2cm}
\centering   
\includegraphics[width=0.56\textwidth,height=3.5in]{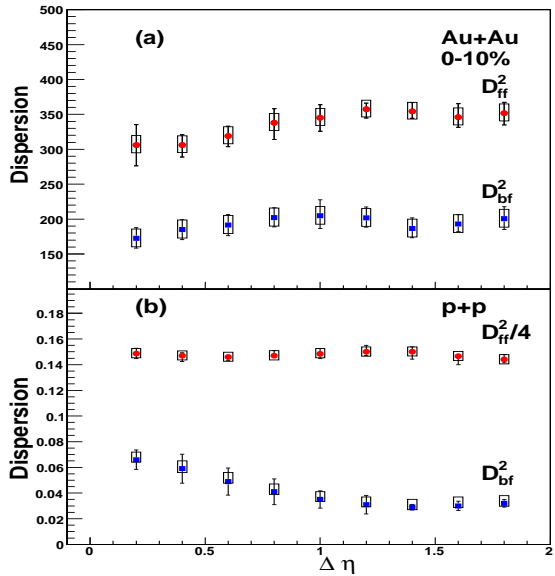}
\vspace*{-1.5cm}    
\caption{(a) Backward-forward dispersion, $D_{bf}^{2}$, and  forward-forward dispersion, $D_{ff}^{2}$, for 0-10\% centrality as a function of $\Delta\eta$ for Au+Au collisions at $\sqrt{s_{NN}}$ = 200 GeV. (b) $D_{bf}^{2}$, and  $D_{ff}^{2}$ for \textit{p+p} collisions at $\sqrt{s}$ = 200 GeV. The error bars represent the systematic point-to-point error. The boxes show the correlated systematic errors.}

\label{PeripAuandpp}
\end{figure}
 
 Short range correlations have been previously observed in high energy hadron-hadron collisions \cite{alner}. The shape of the SRC function is symmetric about midrapidity and has a maximum at $\eta$ = 0. It can be parameterized as $\propto \exp (-\Delta\eta/\lambda)$, where $\lambda$ is the short range correlation length and is found to be $\lambda$ $\sim$ 1. Thus the SRC are significantly reduced
by a separation of $\sim 1.0$ units of pseudorapidity \cite{capela2,larry2}.  
The short range correlation length is smaller in nucleus-nucleus collisions as compared to high energy hadron-hadron collisions \cite{e802,urqmd}.  
 The remaining portion of the correlation strength can be attributed to the LRC. This can be seen in Fig. 1b where the magnitude of the FB correlation strength is zero for 
 $\Delta\eta \sim 1$ for 40-50\% centrality. 
 In case of 0-10\% Au+Au collisions the magnitude of FB correlation strength is 0.6, indicating that 60\% of the observed hadrons are correlated. 
  
     The FB correlation results are compared with the predictions of two models of A+A collisions widely used at RHIC energies - HIJING \cite{hijing} and the Parton String Model (PSM) \cite{amelin2}. The PSM is the Monte Carlo implementation of  the Dual Parton Model (DPM) \cite{capela2} or Quark-Gluon String Model (QGSM) concepts \cite{kaidalov}, considering both soft and semihard components on a partonic level. The HIJING model is based on perturbative QCD processes which lead to multiple jet production and jet interactions in matter \cite{hijing}. Nearly 1 million minimum bias Au+Au collisions at $\sqrt{s_{NN}}$=200 GeV were simulated for each model.
The PSM events were obtained without string fusion options. We used HIJING version 1.383 with default options. We have also simulated 10 million $\textit{p+p}$ minimum bias events at the same cms energy to
 provide the reference for comparison with Au+Au collisions. The correlation analysis was carried out exactly in the same manner as for the data. 
\begin{figure}[hb]
\centering  
\vspace*{-0.3cm}     
\includegraphics[width=0.50\textwidth,height=3.0in]{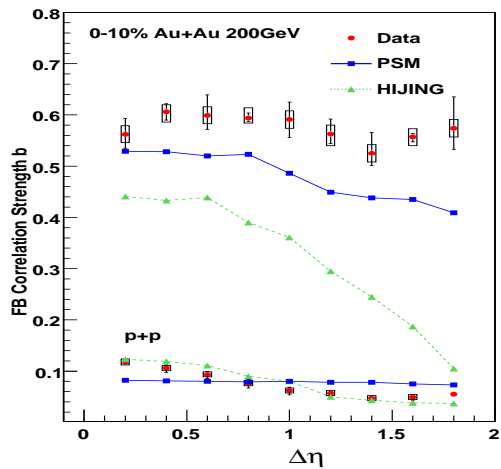}
\vspace*{-1.5cm}
\caption{ The FB correlation strength for 0-10\% most central Au+Au collisions and $\textit{p+p}$ 
from data ( circle), HIJING ( triangle) and PSM ( square). The error bars shown are for data.}
\label{Models}
\end{figure}  
 Both PSM and HIJING predictions are lower than the data as shown in Fig. 3 but PSM exhibits a large LRC for $\Delta\eta > 1.0$ while HIJING predicts significantly smaller correlations than observed in the data. In case of $\textit{p+p}$ collisions the HIJING prediction agrees with the data. The PSM does not show the decrease of $\textit{b}$ with $\Delta\eta$ as seen in the data. 
These trends are illustrated in Fig. 3, where the variation
of the FB correlation strength with $\Delta\eta$ is shown
for Au+Au, HIJING, and PSM. The strong fall of  $\textit{b}$ with $\Delta\eta$ in HIJING provides some constraints on the contribution of impact parameter fluctuations to the correlation strength (Fig.3). Recently, Hadrod-string dynamics (HSD) transport model \cite{hsd}  and CGC \cite{lappi} have addressed the possible effect of impact parameter fluctuations on the correlations with different results.

A description of the FB correlations, which qualitatively agrees with the measured behavior of FB correlation strength is provided by the Dual Parton Model (DPM) \cite{capela1}. 
As mentioned earlier the FB correlation strength is controlled by the  numerator of Eq. \ref{b}. For the case of hadron-hadron collisions:
\begin{eqnarray}
\nonumber D_{bf}^2 = \langle N_{f}N_{b} \rangle - \langle N_{f} \rangle \langle N_{b} \rangle  = \\
\nonumber \langle k \rangle (\langle N_{0f}N_{0b} \rangle - \langle N_{0f} \rangle \langle N_{0b} \rangle ) \\
+ [(\langle k^2 \rangle - \langle k \rangle^2)]\langle N_{0f} \rangle \langle N_{0b} \rangle \hspace{0.05cm}
\label {Dbf}
\end{eqnarray}
where $\langle N_{0f} \rangle$ and $\langle N_{0b} \rangle$ are the average multiplicity of charged particles produced in 
the forward and backward hemispheres in a single elementary inelastic collision \cite{capela2}. The average number of elementary (parton-parton)  inelastic collisions is given by $\langle k \rangle$. 
The first term in Eq. 2 is the correlation between particles produced in the same inelastic collision, representing the SRC in rapidity. The second term, $\langle k^2 \rangle - \langle k \rangle^2$, is due to the fluctuation in the number of elementary inelastic collisions and is controlled by unitarity. This term gives rise to LRC \cite{capela1,capela2}.

 Recently, long range FB multiplicity correlations have also been discussed in the framework of the CGC/glasma motivated phenomenology \cite{larry2,dias}. The glasma provides a QCD based description which includes many features of the DPM approach, in particular the longitudinal rapidity structure \cite{venu}. This model predicts the growth of LRC with collision centrality \cite{larry2}. It has been argued that the long range rapidity correlations are due to the fluctuations of the number of gluons and can only be created at early time shortly after the collision \cite{larry,larry3}.

In summary, this is the first measurement of long-range FB correlation strengths in ultra relativistic nucleus-nucleus collisions. A large long range correlation is observed in central Au+Au collisions that vanishes for 40-50\% centrality. Both DPM and CGC argue that the long range correlations are produced by  multiple parton-parton interactions \cite{capela1,larry}. Multiple parton interactions are necessary for the formation of partonic matter.
It remains an open question whether the DPM and CGC models can describe the LRC reported here and the near-side correlations \cite{tom} simultaneously.
Further studies of the forward-backward correlations using identified baryons and mesons as well as the dependence of the correlations on the collision energy may be able to distinguish between these two models. 
 
 We express our gratitude to C. Pajares and N. Armesto for many fruitful discussions and providing us with the PSM code. We also thank A. Capella, E.G. Ferreiro and Larry McLerran for important discussions. 
We thank the RHIC Operations Group and RCF at BNL, and the NERSC Center 
at LBNL and the resources provided by the Open Science Grid consortium 
for their support. This work was supported in part by the Offices of NP 
and HEP within the U.S. DOE Office of Science, the U.S. NSF, the Sloan 
Foundation, the DFG cluster of excellence `Origin and Structure of the Universe', 
CNRS/IN2P3, RA, RPL, and EMN of France, STFC and EPSRC of the United Kingdom, FAPESP 
of Brazil, the Russian Ministry of Sci. and Tech., the NNSFC, CAS, MoST, 
and MoE of China, IRP and GA of the Czech Republic, FOM of the 
Netherlands, DAE, DST, and CSIR of the Government of India,
the Polish State Committee for Scientific Research,  and the Korea Sci. 
\& Eng. Foundation.



\end{document}